\documentclass{iaus}
\usepackage{graphicx}

\title[IAU 249.~~Astrobiological effects of main-sequence stars]
{Astrobiological effects of F, G, K and M main-sequence stars}

\author[M. Cuntz et al.]   
{M. Cuntz$^1$, L. Gurdemir$^1$, E. F. Guinan$^2$ and R. L. Kurucz$^3$}

\affiliation{
$^1$Department of Physics, University of Texas at Arlington, \\
Arlington, TX 76019-0059, USA \\
email: {\tt cuntz@uta.edu, gurdemir@uta.edu} \\
[\affilskip]
$^2$Department of Astronomy and Astrophysics, Villanova University, \\
Villanova, PA 19085, USA \\
email: {\tt edward.guinan@villanova.edu} \\
[\affilskip]
$^3$Harvard-Smithsonian Center for Astrophysics, \\
Cambridge, MA 02138, USA \\
email: {\tt rkurucz@cfa.harvard.edu}
}

\pubyear{2008}
\volume{xxx}  
\pagerange{119--126}
\jname{Title of your IAU Symposium}
\editors{A.C. Editor, B.D. Editor \& C.E. Editor, eds.}
\begin{document}

\maketitle

\begin{abstract}
We focus on the astrobiological effects of photospheric radiation produced by
main-sequence stars of spectral types F, G, K, and M.  The photospheric radiation
is represented by using realistic spectra, taking into account millions or
hundred of millions of lines for atoms and molecules.  DNA is taken as a proxy
for carbon-based macromolecules, assumed to be the chemical centerpiece of
extraterrestrial life forms.  Emphasis is placed on the investigation of the
radiative environment in conservative as well as generalized habitable zones.
\end{abstract}

\firstsection 

\section{Introduction and Methods}

The centerpiece of all life on Earth is carbon-based biochemistry. It has
repeatedly been surmised that biochemistry based on carbon may also play a
pivotal role in extraterrestrial life forms, if existent.
This is due to the pronounced advantages of carbon, especially compared to its
closest competitor (i.e., silicon), which include: its relatively high
abundance, its bonding properties, and its ability to form very large molecules
as it can combine with hydrogen and other molecules as, e.g., nitrogen and oxygen
in a very large number of ways (\cite[Goldsmith \& Owen 2002]{gol02}).

In the following, we explore the relative damage to carbon-based macromolecules
in the environments of a variety of main-sequence stars using DNA as a proxy
by focussing on the effects of photospheric radiation.
The radiative effects on DNA are considered by applying a DNA action spectrum
(\cite[Horneck 1995]{hor95}) that shows that the damage is strongly wavelength-dependent,
increasing by more than seven orders of magnitude between 400 and 200~nm.
The different regimes are commonly referred to as UV-A, UV-B, and UV-C.
The test planets are assumed to be located in the stellar habitable zone (HZ).
Following the concepts by \cite[Kasting et al. (1993)]{kas93}, we distinguish
between the conservative and generalized HZ.
Stellar photospheric radiation is represented by using realistic spectra
taking into account millions or hundred of millions of lines for
atoms and molecules (\cite[Castelli \& Kurucz 2004]{cas04},
and related publications).  We also consider the effects of attenuation by
an Earth-type planetary atmosphere, which allows us to estimate attenuation
coefficients appropriate to the cases of Earth as today,
Earth 3.5 Gyr ago, and no atmosphere at all (\cite[Cockell 2002]{coc02}).

\begin{figure}
\begin{center}
\includegraphics[width=2.7in]{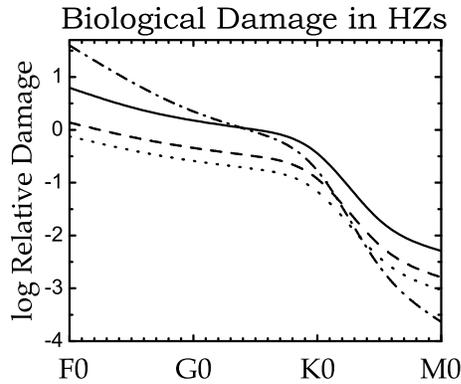}
\caption{
Biological damage to DNA for a planet
at an Earth-equivalent position without an atmosphere
(solid line), an atmosphere akin to Earth 3.5 Gyr ago (dashed line)
and an atmosphere akin to Earth today (dotted line).
The dash-dotted line refers to a planet without an atmosphere at
a distance of 1~AU.
}
\end{center}
\end{figure}

\begin{figure}
\begin{center}
\includegraphics[width=2.7in]{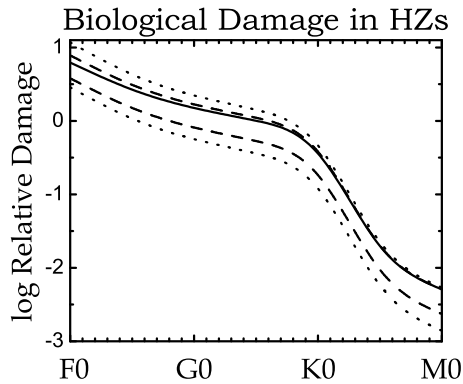}
\caption{Biological damage to DNA for a planet (no atmosphere)
at an Earth-equivalent position (solid line), at the limits of the
conservative HZ (dashed lines) and at the limits of the generalized HZ
(dotted lines).
}
\end{center}
\end{figure}

\begin{figure}
\begin{center}
\includegraphics[width=3.3in]{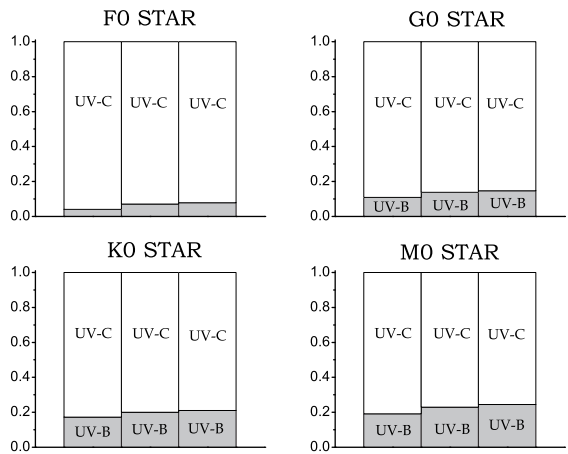}
\caption{Relative significance of UV-A, UV-B, and UV-C for the
damage to DNA for a planet without an atmosphere
(left), an atmosphere akin to Earth 3.5 Gyr ago (center)
and an atmosphere akin to Earth today (right) for different types
of main-sequence stars.  Note that the fraction due
to UV-A is unidentifiable.
}
\end{center}
\end{figure}

\section{Results and Conclusions}

Our results are presented in Figs. 1, 2, and 3.  The first two figures show
the relative damage to DNA due to stars between spectral type F0 and M0,
normalized to today's Earth.  We also considered planets at the inner and outer
edge of either the conservative or generalized HZ as well as planets of different
atmospheric attenuation.

Based on our studies we arrive at the following conclusions:
(1) All main-sequence stars of spectral type F to M have the potential of
damaging DNA due to UV radiation.  The amount of damage strongly depends
on the stellar spectral type, the type of the planetary atmosphere and
the position of the planet in the habitable zone (HZ); see 
\cite[Cockell (1999)]{coc99} for previous results.
(2) The damage to DNA for a planet in the HZ around an F-star (Earth-equivalent
distance) due to photospheric radiation is significantly higher (factor 5) compared
to planet Earth around the Sun, which in turn is significantly higher than for an
Earth-equivalent planet around an M-star (factor 180).
(3) We also found that the damage is most severe in
the case of no atmosphere at all, somewhat less severe for an atmosphere
corresponding to Earth 3.5 Gyr ago, and least severe for an atmosphere like
Earth today.
(4) Any damage due to photospheric stellar radiation is mostly due to UV-C.
The relative importance of UV-B is between 5\% (F-stars) and 20\% (M-stars).
Note that damage due to UV-A is virtually nonexistent (see Fig. 3).

Our results are of general interest for the future search of planets
in stellar HZ (e.g., \cite[Turnbull \& Tarter 2003]{tur03}).
They also reinforce the notion
that habitability may at least in principle be possible around M-type stars, as
previously discussed by \cite[Tarter et al. (2007)]{tar07}.  Note however that
a more detailed analysis also requires the consideration of chromospheric
UV radiation, especially flares (e.g., \cite[Robinson et al. 2005]{rob05}), as
well as the detailed treatment of planetary atmospheric photochemistry, including
the build-up and destruction of ozone, as pointed out by 
\cite[Segura et al. (2003, 2005)]{seg03,seg05} and others.

\end{document}